\documentclass[preprint,showpacs,preprintnumbers,amssymb,aps,nofootinbib]{revtex4}
\usepackage{amsfonts,color,graphicx,epsfig,graphicx,amsmath,multirow,slashed}

\begin{document}

\title{Studies of $Z$ boson decay into double $\Upsilon$ mesons at the NLO QCD accuracy}

\author{Cong Li}
\author{Ying-Zhao Jiang}
\author{Zhan Sun}
\email{sunzhan_hep@163.com}

\affiliation{
\footnotesize
Department of Physics, Guizhou Minzu University, Guiyang 550025, People's Republic of China. \\
}

\date{\today}

\begin{abstract}

In this paper, we employ the nonrelativistic QCD factorization to conduct a comprehensive examination of the $Z$ boson decay into a pair of $\Upsilon$ mesons, achieving accuracy at the next-to-leading-order (NLO) in $\alpha_s$. Our calculations demonstrate that the QED diagrams are indispensable in comparison to the pure QCD diagrams, and the implementation of QCD corrections markedly enhance the QCD results, whereas it substantially diminish the QED results. To ensure consistency with the experimental methodology, we have taken into account the feed-down transitions originating from higher excited states, which exhibit significant relevance. Combining all the contributions, we arrive at the NLO prediction of $\mathcal{B}_{Z \to \Upsilon(nS)+\Upsilon(mS)} \sim 10^{-11}$, which is notably lower than the upper limits set by CMS.

\pacs{12.38.Bx, 12.39.Jh, 14.40.Pq}

\end{abstract}

\maketitle

\section{Introduction}\label{intro}

Heavy-quarkonium production in $Z$ boson decay has garnered significant interest from both theoretical and experimental physicists over the past decades \cite{z decay 1,z decay 2,z decay 3,z decay 4,z decay 5,z decay 6,z decay 7,z decay 8,z decay 9,z decay 10,z decay 11,z decay 12,z decay 13,z decay 14,z decay 15,z decay 16,z decay 17,z decay 18,z decay 19,z decay 20,z decay 21,z decay 22,z decay 23,z decay 24,z decay 25,z decay 26,z decay 27,z decay 28,z decay 29,z decay 30,z decay 31,z decay 32,z decay 33,z decay 34,z decay 35,z decay 36,z decay 37,z decay 38,z decay 39,z decay 40,z decay 41,z decay 42,z decay 43}. In the various decay channels of the $Z$ boson into heavy quarkonium, the yield of double $J/\psi$ (or $\Upsilon$) mesons holds a notable advantage due to the unique experimental signature arising from their subsequent decay into four muons, which can distinctly be recognized and analyzed.

In 2019, the CMS Collaboration measured the branching ratio of the $Z$ boson decaying to double $J/\psi$ \cite{CMS:2019wch}, denoted as $\mathcal{B}_{Z \to J/\psi+J/\psi} < 2.2 \times 10^{-6}$. This upper limit was recently updated to be $1.4 \times 10^{-6}$ by analyzing a larger sample of data \cite{CMS:2022fsq}. At leading order (LO) in $\alpha_s$, the standard model provides a prediction of $\mathcal{B}_{Z \to J/\psi+J/\psi}\sim 10^{-12}$ \cite{Likhoded:2017jmx}, which was upgraded to $10^{-10}$ by further evaluating the QED contributions resulting from the virtual-photon effects \cite{Gao:2022mwa}. Recently, Li $et~al.$ suggested that the QCD corrections play a crucial role in enhancing the contributions from QCD diagrams and simultaneously diminishing the contributions from QED diagrams, ultimately maintaining the next-to-leading-order (NLO) results at the order of $10^{-10}$ \cite{z decay 40}.  

In addition to the double $J/\psi$ yield, the CMS group has also explored the Z boson decaying into a pair of $\Upsilon$ mesons, thereby establishing upper limits on the branching ratio \cite{CMS:2019wch,CMS:2022fsq}
\begin{eqnarray}
\mathcal{B}_{Z\to\Upsilon(mS)+\Upsilon(nS)} \leq 3.9\times10^{-7}, \nonumber \\
\mathcal{B}_{Z\to\Upsilon(1S)+\Upsilon(1S)} \leq 1.8\times10^{-6}.
\end{eqnarray}
By taking into account both the QCD and QED diagrams, Gao $et~al.$ provided an estimation of $\mathcal{B}_{Z \to \Upsilon(1S)+\Upsilon(1S)}$ at the LO QCD accuracy \cite{Gao:2022mwa}. In light of the considerable effect that NLO QCD corrections have on double-charmonia production in $e^{-}e^{+}$ annihilation \cite{{Zhang:2005cha,Gong:2007db,Zhang:2008gp,Brambilla:2010cs,Dong:2011fb,Sun:2018rgx,Sun:2021tma}}, it is prudent to investigate whether high-order terms in $\alpha_s$ could similarly engender a significant boost in the yield of double $\Upsilon$ mesons in $Z$ boson decay. To achieve this, in the present work, we will examine the decay of $Z$ boson into $\Upsilon$ pair with NLO precision in $\alpha_s$, incorporating both QCD and QED diagrams within the nonrelativistic QCD (NRQCD) formalism \cite{NRQCD1}. Note that, the measured branching ratio of $Z$ boson decay into double $\Upsilon$ mesons includes considerations for feed-down transitions. Consequently, this evaluation will also consider the impact of higher excited states.

It is noteworthy that, the considerable mass of the $b$ quark typically leads to a more rapidly converging perturbative series in the expansion of $\alpha_s$ and $v^2$, compared to the case of the $c$ quark. Coupled with the significant production rates of Z bosons at the LHC, the proposed HL-LHC, or the future CEPC \cite{CEPC}, the yields of double $\Upsilon$ mesons in $Z$ decay would offer an excellent laboratory for probing the mechanisms governing heavy quarkonium formation.
 
The remainder of this paper is structured as follows: Section \ref{cal} provides an outline of the calculation formalism. This is followed by the presentation of phenomenological results and discussions in Section \ref{results}. Finally, Section \ref{sum} is dedicated to a summary.

\section{Calculation formalism}\label{cal}
\subsection{Theoretical Framework}
As previously discussed, the CMS measurements have incorporated feed-down transitions; consequently, our calculations will assess the contributions stemming from higher excited states, including those of the $\Upsilon(mS)$ and $\chi_{b}(nP)$ mesons. Within the NRQCD framework \cite{NRQCD1,NRQCD2}, the decay width of $Z \to H_{1}(b\bar{b})+H_{2}(b\bar{b})$ can be factorized as
\begin{eqnarray}
\Gamma=\hat{\Gamma}_{Z \to b\bar{b}[n_1]+b\bar{b}[n_2]}\langle \mathcal{O}^{H_1}(n_1)\rangle \langle \mathcal{O}^{H_2}(n_2)\rangle,\label{eq3}
\end{eqnarray}
where $\hat{\Gamma}_{Z \to b\bar{b}[n_1]+b\bar{b}[n_2]}$ represents the perturbative calculable short-distance coefficients (SDCs), denoting the production of the intermediate states consisting of $b\bar{b}[n_1]$ and $b\bar{b}[n_2]$. With the restriction to color-singlet contributions, we have $n_{1}=^3S_1^{1}$ and $n_2=^3S_1^{1}(\textrm{or}~^3P_J^{1})$ ($J=0,1,2$).\footnote{When addressing the feed-down transitions, we exclude the evaluation of $Z \to \chi_{b}(nP)+\chi_{b}(n'P)$. This process is anticipated to contribute $Z \to \Upsilon(mS)+\Upsilon(m'S)$ through a multi-step decay, which involves at least two decay stages.} The universal nonperturbative long distant matrix element (LDME) $\langle \mathcal{O}^{H_{1(2)}}(n_{1(2)})\rangle$ stands for the probabilities of transitions from $b\bar{b}[n_{1(2)}]$ into the $H_{1(2)}$ meson states.

The $\hat{\Gamma}_{Z \to b\bar{b}[n_1]+b\bar{b}[n_2]}$ can further be expressed as
\begin{eqnarray}
\hat{\Gamma}_{Z \to b\bar{b}[n_1]+b\bar{b}[n_2]}=\frac{\kappa}{2m_Z}\frac{1}{N_I N_s}|\mathcal{M}|^2,\label{eq4}
\end{eqnarray}
where $|\mathcal{M}|^2$ is the squared matrix elements, $1/N_s$ is the spin average factor of the initial $Z$ boson multiplied by the identity factor $N_I$ of the two final $b\bar{b}$ states. For $b\bar{b}[n_1]$ and $b\bar{b}[n_2]$ representing the same states, $N_I$ is set to $1/2!$; however, if they are distinct states, $N_I$ equals 1. The symbol $\kappa$ denotes the factor originating from the standard two-body phase space.

Based on the framework used to handle $Z-J/\psi+J/\psi$ \cite{z decay 40}, we include terms up to the $\alpha^3$ order and achieve NLO accuracy in $\alpha_s$. The squared matrix elements specified in Eq. (\ref{eq4}) can subsequently be expressed as follows
\begin{eqnarray}
&&\bigg|\left(\mathcal{M}_{\alpha^{\frac{1}{2}}\alpha_{s}}+\mathcal{M}_{\alpha^{\frac{1}{2}}\alpha^2_{s}}\right)+\left(\mathcal{M}_{\alpha^{\frac{3}{2}}}+\mathcal{M}_{\alpha^{\frac{3}{2}}\alpha_{s}}\right)\bigg|^2 \nonumber \\
&=&\big| \mathcal{M}_{\alpha^{\frac{1}{2}}\alpha_{s}} \big|^2+2 \textrm{Re}\left( \mathcal{M}^{*}_{\alpha^{\frac{1}{2}}\alpha_{s}} \mathcal{M}_{\alpha^{\frac{1}{2}}\alpha^2_{s}} \right) \nonumber \\
&& + 2 \textrm{Re}\left( \mathcal{M}^{*}_{\alpha^{\frac{1}{2}}\alpha_{s}} \mathcal{M}_{\alpha^{\frac{3}{2}}} \right)+2 \textrm{Re}\left( \mathcal{M}^{*}_{\alpha^{\frac{1}{2}}\alpha_{s}} \mathcal{M}_{\alpha^{\frac{3}{2}}\alpha_{s}}+\mathcal{M}^{*}_{\alpha^{\frac{3}{2}}} \mathcal{M}_{\alpha^{\frac{1}{2}}\alpha^2_{s}} \right) \nonumber \\
&& + \big| \mathcal{M}_{\alpha^{\frac{3}{2}}} \big|^2+2 \textrm{Re}\left( \mathcal{M}^{*}_{\alpha^{\frac{3}{2}}} \mathcal{M}_{\alpha^{\frac{3}{2}}\alpha_{s}} \right) + \cdots. \label{eq5}
\end{eqnarray}
Accordingly, we decompose the SDCs into three distinct components,
\begin{eqnarray}
\hat{\Gamma} = \hat{\Gamma}^{(0,1)}_{1}+\hat{\Gamma}^{(0,1)}_{2}+\hat{\Gamma}^{(0,1)}_{3},\label{eq6}
\end{eqnarray}
where
\begin{eqnarray}
\hat{\Gamma}^{(0)}_{1}&\propto&\big| \mathcal{M}_{\alpha^{\frac{1}{2}}\alpha_{s}} \big|^2, \nonumber \\
\hat{\Gamma}^{(1)}_{1}&\propto&2 \textrm{Re}\left( \mathcal{M}^{*}_{\alpha^{\frac{1}{2}}\alpha_{s}} \mathcal{M}_{\alpha^{\frac{1}{2}}\alpha^2_{s}} \right), \nonumber \\
\hat{\Gamma}^{(0)}_{2}&\propto&2 \textrm{Re}\left( \mathcal{M}^{*}_{\alpha^{\frac{1}{2}}\alpha_{s}} \mathcal{M}_{\alpha^{\frac{3}{2}}} \right), \nonumber \\
\hat{\Gamma}^{(1)}_{2}&\propto& 2 \textrm{Re}\left( \mathcal{M}^{*}_{\alpha^{\frac{1}{2}}\alpha_{s}} \mathcal{M}_{\alpha^{\frac{3}{2}}\alpha_{s}}+\mathcal{M}^{*}_{\alpha^{\frac{3}{2}}} \mathcal{M}_{\alpha^{\frac{1}{2}}\alpha^2_{s}} \right), \nonumber \\
\hat{\Gamma}^{(0)}_{3}&\propto&\big| \mathcal{M}_{\alpha^{\frac{3}{2}}} \big|^2, \nonumber \\
\hat{\Gamma}^{(1)}_{3}&\propto&2 \textrm{Re}\left( \mathcal{M}^{*}_{\alpha^{\frac{3}{2}}} \mathcal{M}_{\alpha^{\frac{3}{2}}\alpha_{s}} \right).
\label{eq7}
\end{eqnarray}
The subscripts 1, 2, and 3 represent the order in $\alpha$, while the superscript $0(1)$ denotes the terms at the LO (or NLO) level in $ \alpha_s $.
\begin{figure}[!h]
\begin{center}
\hspace{0cm}\includegraphics[width=0.65\textwidth]{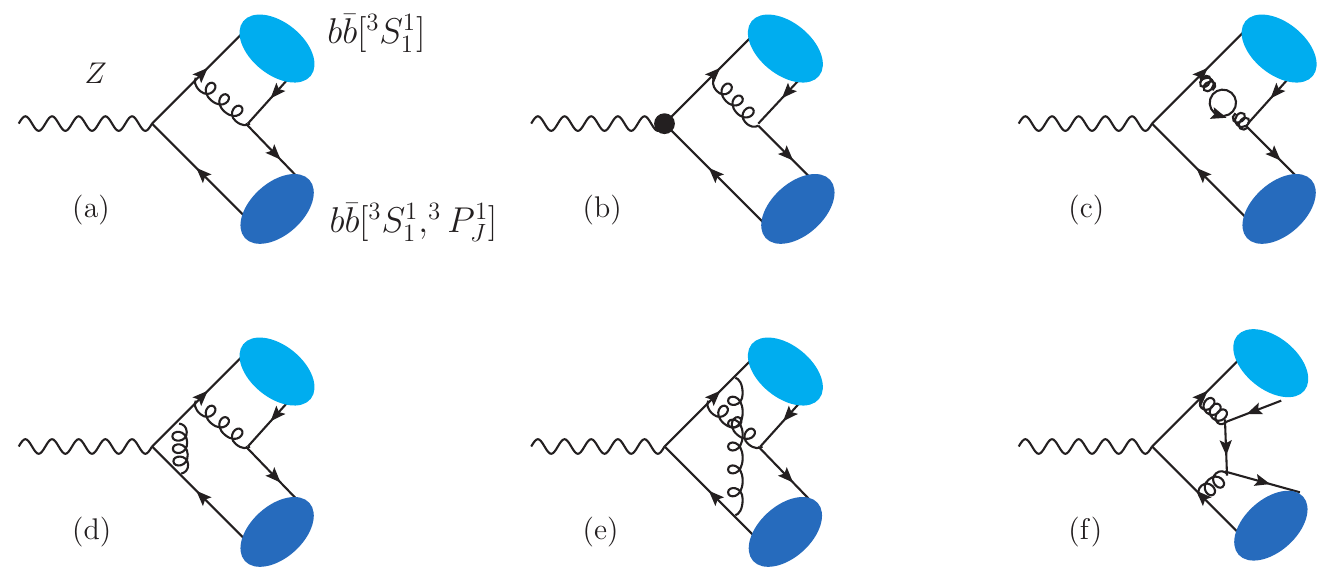}
\caption{\label{fig:QCD Feynman Diagrams}
Representative QCD Feynman diagrams for $Z \to b\bar{b}[^3S_1^{1}]+b\bar{b}[^3S_1^{1},^3P_J^{1}]$. Diagram (a) (${\alpha^{\frac{1}{2}}\alpha_{s}}$ order) is the QCD tree-level diagram. Diagrams (b)-(f) $({\alpha^{\frac{1}{2}}\alpha^2_{s}}~\textrm{order})$ depict the NLO QCD corrections to (a). Diagram (b) specifically represents the counter-term diagram.}
\end{center}
\end{figure}

\begin{figure}[!h]
\begin{center}
\hspace{0cm}\includegraphics[width=0.995\textwidth]{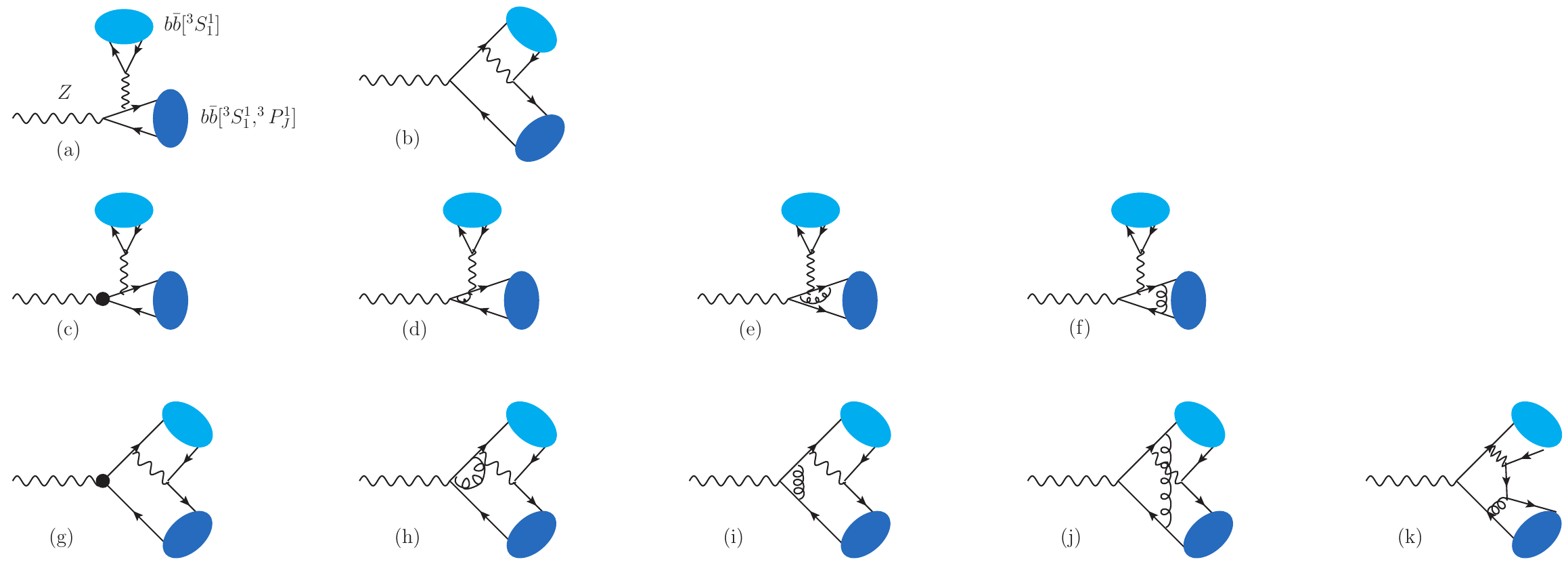}
\caption{\label{fig:QED Feynman Diagrams}
Representative QED Feynman diagrams for $Z \to b\bar{b}[^3S_1^{1}]+b\bar{b}[^3S_1^{1},^3P_J^{1}]$. Diagrams (a,b) $(\alpha^{\frac{3}{2}}$ order) are the QCD tree-level diagrams. Diagrams (c)-(k) ($\alpha^{\frac{3}{2}}\alpha_{s}$ order) depict the NLO QCD corrections to (a,b). Diagrams (c,g) specifically represent the counter-term diagram.}
\end{center}
\end{figure}

The representative Feynman diagrams of $\mathcal{M}_{\alpha^{\frac{1}{2}}\alpha_{s}}$, $\mathcal{M}_{\alpha^{\frac{1}{2}}\alpha^2_{s}}$, $\mathcal{M}_{\alpha^{\frac{3}{2}}}$, and $\mathcal{M}_{\alpha^{\frac{3}{2}}\alpha_{s}}$ are displayed in Figs. \ref{fig:QCD Feynman Diagrams} and \ref{fig:QED Feynman Diagrams}. Fig. \ref{fig:QCD Feynman Diagrams}(a) ($\alpha^{\frac{1}{2}}\alpha_{s}$ order) demonstrates the QCD tree-level diagram (4 diagrams), whereas Figs. \ref{fig:QCD Feynman Diagrams}(b)-(f) elucidate the subsequent QCD NLO corrections, which comprise 56 individual one-loop diagrams and 20 counter-term diagrams. In Figs. \ref{fig:QED Feynman Diagrams}(a,b) ($\alpha^{\frac{3}{2}}$ order) we observe the tree-level diagrams pertinent to QED, encompassing 8 diagrams in total. Furthermore, Figs. \ref{fig:QED Feynman Diagrams}(c)-(k) depict the high-order corrections in $\alpha_s$, including 52 one-loop diagrams and 32 counter-term diagrams.

\subsection{The decay width}
According to Eqs. (\ref{eq3}) and (\ref{eq4}), the decay width can generally be written as
\begin{eqnarray}
\Gamma^{\textrm{(0,1)}}_{i}=\frac{\kappa}{2m_Z}\frac{1}{N_I N_s} \times \zeta_{i} \times \mathcal{C}^{(0,1)}_{i} \times \langle \mathcal O^{H_1}(^3S_{1}^{1}) \rangle\langle \mathcal O^{H_2}(^3S_{1}^{1},^3P_{J}^{1}) \rangle,~i=1,2,3,
\label{eq8}
\end{eqnarray}
where $\kappa$ is equal to $\frac{\sqrt{m^2_Z-16 m^2_b}}{8 \pi m_Z }$ and $N_s=3$.

For the yield of $b\bar{b}[^3S_1^{1}]+b\bar{b}[^3S_1^{1}]$, the $\zeta_{i}$ takes the following form ($e_b=-\frac{1}{3}$)
\begin{eqnarray}
\zeta_{i}&=&\frac{\pi^3}{16 m_b^6 \sin^2\theta_{\textrm{w}} \cos^2\theta_{\textrm{w}}} e^{2(i-1)}_b \alpha^{i}\alpha^{3-i}_s,
\label{eq9}
\end{eqnarray}
while for the scenario involving $b\bar{b}[^3S_1^{1}]+b\bar{b}[^3P_J^{1}]$,
\begin{eqnarray}
\zeta_{i}&=&\frac{\pi^3 (4\sin^2\theta_{\textrm{w}}-3)^2}{144 m_b^6 \sin^2\theta_{\textrm{w}} \cos^2\theta_{\textrm{w}}} e^{2(i-1)}_b \alpha^{i}\alpha^{3-i}_s.
\label{eq10}
\end{eqnarray}

The LDMEs $\langle \mathcal O^{H}(^3S_{1}^{1},^3P_{J}^{1}) \rangle$ can be expressed in terms of the wave functions at the origin by utilizing the following formulae
\begin{eqnarray}
\langle \mathcal O^{\Upsilon(nS)}(^3S_1^{1}) \rangle &=& \frac{1}{4\pi}|R_{nS}(0)|^2, \nonumber \\
\langle \mathcal O^{\chi_{bJ}(mP)}(^3P_J^{1}) \rangle &=& \frac{3}{4\pi}|R'_{mP}(0)|^2.
\label{eq11}
\end{eqnarray}

\subsubsection{LO}

The coefficients of $\mathcal{C}^{(0)}_{i}$ derived from the LO processes, as illustrated in Figs. \ref{fig:QCD Feynman Diagrams}(a) and \ref{fig:QED Feynman Diagrams}(a,b) are free from divergences. In the following, we provide the expressions for various processes ($r\equiv\frac{m^2_Z}{4m_b^2}$).
\begin{itemize}
\item[1)]
$b\bar{b}[^3S_1^{1}]+b\bar{b}[^3S_1^{1}]$
\begin{eqnarray}
&&\mathcal{C}^{(0)}_{1}=\frac{65536 m_b^2 ( r^2 -10 r+24)}{9r^4}, \nonumber \\
&&\mathcal{C}^{(0)}_{2}=\frac{16384 m_b^2 (3r+2) ( r^2 -10 r+24)}{3r^4}, \nonumber \\
&&\mathcal{C}^{(0)}_{3}=\frac{1024 m_b^2 (3r+2)^2 ( r^2 -10 r+24)}{r^4}.
\label{eq12}
\end{eqnarray}
\item[2)]
$b\bar{b}[^3S_1^{1}]+b\bar{b}[^3P_J^{0}]$
\begin{eqnarray}
&&\mathcal {C}^{(0)}_{1}= \frac{16384(r^{4}+182r^{3}-428r^{2}+152r+144)}{27r^5},\nonumber \\
&&\mathcal {C}^{(0)}_{2}=\frac{32768(7r^{4}-7r^{3}-59r^{2}+56r+36)}{9r^5},\nonumber\\
&&\mathcal {C}^{(0)}_{3}=\frac{512(9r^{5}+2r^{4}-152r^{3}-16r^{2}+592r+288)}{3r^5}.
\label{eq13}
\end{eqnarray}
\item[3)]
$b\bar{b}[^3S_1^{1}]+b\bar{b}[^3P_J^{1}]$
\begin{eqnarray}
&&\mathcal {C}^{(0)}_{1}= \frac{131072(2r^{3}-12r^{2}+13r+18)}{9r^5},\nonumber \\
&&\mathcal {C}^{(0)}_{2}= -\frac{16384(19r^{3}-24r^{2}-88r-72)}{3r^5},\nonumber \\
&&\mathcal {C}^{(0)}_{3}= \frac{1024(9r^{5}-9r^{4}+34r^{3}+228r^{2}+248r+144)}{r^5}.
\label{eq14}
\end{eqnarray}
\item[4)]
$b\bar{b}[^3S_1^{1}]+b\bar{b}[^3P_J^{2}]$
\begin{eqnarray}
&&\mathcal {C}^{(0)}_{1}= \frac{32768(r^{4}+20r^{3}-188r^{2}+308r+360)}{27r^5},\nonumber \\
&&\mathcal {C}^{(0)}_{2}= \frac{16384(10r^{4}-37r^{3}-104r^{2}+488r+360)}{9r^5}, \nonumber \\
&&\mathcal {C}^{(0)}_{3}= \frac{1024(9r^{5}+11r^{4}-134r^{3}+140r^{2}+1336r+720)}{3r^5}.
\label{eq15}
\end{eqnarray}
\end{itemize}

\subsubsection{NLO}

We utilize the dimensional regularization with $D=4-2\epsilon$ to isolate the ultraviolet (UV) and infrared (IR) divergences. The on-mass-shell (OS) scheme is employed to set the renormalization constants for the $b$-quark mass ($Z_m$) and heavy-quark filed ($Z_2$); the minimal-subtraction ($\overline{MS}$) scheme is adopted for the QCD-gauge coupling ($Z_g$) and the gluon filed $Z_3$. The renormalization constants are taken as
\begin{eqnarray}
\delta Z_{m}^{OS}&=& -3 C_{F} \frac{\alpha_s}{4\pi}N_{\epsilon}\left[\frac{1}{\epsilon_{\textrm{UV}}}+\frac{4}{3}+2\textrm{ln}{2}\right], \nonumber \\
\delta Z_{2}^{OS}&=& - C_{F} \frac{\alpha_s}{4\pi}N_{\epsilon}\left[\frac{1}{\epsilon_{\textrm{UV}}}+\frac{2}{\epsilon_{\textrm{IR}}}+4+6 \textrm{ln}{2}\right], \nonumber \\
\delta Z_{3}^{\overline{MS}}&=& \frac{\alpha_s }{4\pi}(\beta_{0}-2 C_{A})N_{\epsilon}\left[\frac{1}{\epsilon_{\textrm{UV}}}+\textrm{ln}\frac{4m_b^2}{\mu_r^2}\right], \nonumber \\
\delta Z_{g}^{\overline{MS}}&=& -\frac{\beta_{0}}{2}\frac{\alpha_s }{4\pi}N_{\epsilon}\left[\frac{1} {\epsilon_{\textrm{UV}}}+\textrm{ln}\frac{4m_b^2}{\mu_r^2}\right], \label{eq16}
\end{eqnarray}
where $N_{\epsilon}= \frac{1}{\Gamma[1-\epsilon]}\left(\frac{4\pi\mu_r^2}{4m_b^2}\right)^{\epsilon}$ is an overall factor, $\gamma_E$ is the Euler's constant, and $\beta_{0}=\frac{11}{3}C_A-\frac{4}{3}T_Fn_f$ is the one-loop coefficient of the $\beta$ function. $n_f(=n_{L}+n_{H})$ represents the number of the active-quark flavors; $n_{L}(=4)$ and $n_{H}(=1)$ denote the number of the light- and heavy-quark flavors, respectively.\footnote{In our calculations, the $c$ quark is considered a light quark in comparison to the $b$ quark; therefore, we assign $n_l=4$ for the number of light quarks (encompassing $u, d, s$, and $c$) and $n_h=1$ for the number of heavy quarks (specifically, the $b$ quark).} In ${\rm SU}(3)$, the color factors are given by $T_F=\frac{1}{2}$, $C_F=\frac{4}{3}$, and $C_A=3$.

Incorporating the QCD corrections allows us to derive the coefficients for $\mathcal{C}^{(1)}_{i}$, which can be formulated in a generic manner
\begin{eqnarray}
\mathcal{C}^{(0)}_{i}+\mathcal{C}^{(1)}_{i}=\mathcal{C}^{(0)}_{i} \left[ 1+\frac{\alpha_s}{\pi} \left(\xi_{i}\beta_{0}\textrm{ln}\frac{\mu_r^2}{4m_b^2}+a_{i} n_{L}+b_{i} n_{H}+c_{i} \right) \right],\label{eq17}
\end{eqnarray}
where $\xi_1=\frac{1}{2}$, $\xi_2=\frac{1}{4}$, and $\xi_3=0$. The coefficients $a_{i}$, $b_{i}$, and $c_{i}$ are dependent solely on the variables of $r$ and $m_b$. Their fully analytical expressions can be found in Appendices \ref{A1}-\ref{A24}.

To generate all the necessary Feynman diagrams and corresponding analytical amplitudes, we use the \texttt{FeynArts} package \cite{Hahn:2000kx}. We then employ the \texttt{FeynCalc} package \cite{Mertig:1990an} to handle the traces of the $\gamma$ and color matrices, which transforms the hard scattering amplitudes into expressions with loop integrals. When calculating the $D$-dimensional $\gamma$ traces that incorporate a single $\gamma_{5}$ matrix and involve UV and/or IR divergences, we follow the scheme outlined in Refs. \cite{Korner:1991sx,z decay 4, z decay 22} and choose the same starting point ($Z$-vertex) to write down the amplitudes without implementation of cyclicity. Subsequently, we utilize our self-written $\textit{Mathematica}$ codes that include implementations of \texttt{Apart} \cite{Feng:2012iq} and \texttt{FIRE} \cite{Smirnov:2008iw} to reduce these loop integrals to a set of irreducible Master Integrals (MIs). The fully-analytical expressions for these MIs can be found in Appendix \ref{MI1}-\ref{MI12}. As a cross check, we simultaneously adopt the \texttt{LoopTools} \cite{Hahn:1998yk} package to numerically evaluate these MIs, obtaining the same numerical results.

\section{Phenomenological results}\label{results}

The parameters incorporated into our calculations are defined as follows: $m_b=4.7 \pm 0.1$ GeV, $m_Z=91.1876$ GeV, and $\alpha=1/128$. Additionally, we employ the two-loop $\alpha_s$ running coupling constant in our analysis.

The wave functions in Eq. (\ref{eq11}) read \cite{Eichten:1995ch}
\begin{eqnarray}
|R_{1S}(0)|^2&=&6.477~\textrm{GeV}^3,~~~~~~|R_{2S}(0)|^2=3.234~\textrm{GeV}^3,\nonumber \\
|R_{3S}(0)|^2&=&2.474~\textrm{GeV}^3, \nonumber \\
|R^{'}_{1P}(0)|^2&=&1.417~\textrm{GeV}^5,~~~~~~|R^{'}_{2P}(0)|^2=1.653~\textrm{GeV}^5,\nonumber \\
|R^{'}_{3P}(0)|^2&=&1.794~\textrm{GeV}^5. \label{eq18}
\end{eqnarray}

The branching ratios of diverse excited bottomonium states transitioning to lower energy states are documented in Refs. \cite{Br1,Br2,Br3,PDG}.

\begin{table*}[htb]
\begin{center}
\caption{SDCs (in unit: $10^{-12}$ $\textrm{GeV}^{-5}$) corresponding to the process $Z \to b\bar{b}[^3S_1^{1}]+b\bar{b}[^3S_1^{1}]$ with $m_b=4.7$ GeV and $\mu_r=\frac{m_Z}{2}$. $\hat{\Gamma}^{\textrm{LO}}_{\textrm{total}}$ incorporates the individual contributions of $\hat{\Gamma}_{1}^{(0)}$, $\hat{\Gamma}_{2}^{(0)}$, and $\hat{\Gamma}_{3}^{(0)}$, while $\hat{\Gamma}^{\textrm{NLO}}_{\textrm{total}}$ refers to the combined sum of the contributions of $\hat{\Gamma}_{1,2,3}^{(0,1)}$.}
\label{tab: 1}
\begin{tabular}{|cc|cc|cc|cc|cc|cccc}
\hline
$\hat\Gamma^{(0)}_{1}$ & $\hat\Gamma^{(1)}_{1}$ & $\hat\Gamma^{(0)}_{2}$ & $\hat\Gamma^{(1)}_{2}$ & $\hat\Gamma^{(0)}_{3}$ & $\hat\Gamma^{(1)}_{3}$ & $\hat\Gamma^{\textrm{LO}}_{\textrm{total}}$ & $\hat\Gamma^{\textrm{NLO}}_{\textrm{total}}$\\ \hline
$165.2$ & $244.9$ & $229.4$ & $134.4$ & $79.67$ & $-24.79$ & $474.3$ & $828.8$\\
\hline
\end{tabular}
\end{center}
\end{table*}

\begin{table*}[htb]
\begin{center}
\caption{SDCs (in unit: $10^{-12}$ $\textrm{GeV}^{-7}$) corresponding to the process $Z \to b\bar{b}[^3S_1^{1}]+b\bar{b}[^3P_J^{1}]$ with $m_b=4.7$ GeV and $\mu_r=\frac{m_Z}{2}$. $\hat{\Gamma}^{\textrm{LO}}_{\textrm{total}}$ incorporates the individual contributions of $\hat{\Gamma}_{1}^{(0)}$, $\hat{\Gamma}_{2}^{(0)}$, and $\hat{\Gamma}_{3}^{(0)}$, while $\hat{\Gamma}^{\textrm{NLO}}_{\textrm{total}}$ refers to the combined sum of the contributions of $\hat{\Gamma}_{1,2,3}^{(0,1)}$.}
\label{tab: 2}
\begin{tabular}{|c|cc|cc|cc|cc|cccc}
\hline
$~$ & $\hat\Gamma^{(0)}_{1}$ & $\hat\Gamma^{(1)}_{1}$ & $\hat\Gamma^{(0)}_{2}$ & $\hat\Gamma^{(1)}_{2}$ & $\hat\Gamma^{(0)}_{3}$ & $\hat\Gamma^{(1)}_{3}$ & $\hat\Gamma^{\textrm{LO}}_{\textrm{total}}$ & $\hat\Gamma^{\textrm{NLO}}_{\textrm{total}}$\\ \hline
$J=0$ & $92.12$ & $81.42$ & $8.631$ & $3.358$ & $0.323$ & $-0.070$ & $101.1$ & $185.8$\\
$J=1$ & $15.26$ & $10.84$ & $-0.373$ & $2.282$ & $1.916$ & $-0.499$ & $16.80$ & $29.42$\\
$J=2$ & $76.08$ & $15.61$ & $5.985$ & $-0.700$ & $0.652$ & $-0.413$ & $82.72$ & $97.21$\\
\hline
\end{tabular}
\end{center}
\end{table*}

Before proceeding further, let us first scrutinize the SDCs. Inspecting the data in Tables \ref{tab: 1} and \ref{tab: 2}, we find
\begin{itemize}
\item[1)]
In the production of double $b\bar{b}[^3S_1^{1}]$ states, considering the LO level in $\alpha_s$, the inclusion of $\hat{\Gamma}_{2,3}^{(0)}$ enhances the pure QCD results, denoted as $\hat{\Gamma}_{1}^{(0)}$, by approximately a factor of 2. When QCD corrections are incorporated, the QCD results are further augmented by about 1.5 times, as indicated by the ratio $\hat{\Gamma}_{1}^{(1)}/\hat{\Gamma}_{1}^{(0)}$. However, high-order terms in $\alpha_s$ significantly reduce the QED results, with $(\hat{\Gamma}_{3}^{(1)}+\hat{\Gamma}_{3}^{(0)})/\hat{\Gamma}_{3}^{(0)}$ being approximately $70\%$. Due to the interference between QCD and QED results, $\hat{\Gamma}_{2}^{(0)}$ experience a $40\%$ enhancement when QCD corrections are considered, as reflected in the ratio $\hat{\Gamma}_{2}^{(1)}/\hat{\Gamma}_{2}^{(0)}$.
\item[2)]
Regarding the $b\bar{b}[^3S_1^{1}]$ yield associated with $b\bar{b}[^3P_J^{1}]$, the LO SDCs $\hat{\Gamma}_{2,3}^{(0)}$ amplify $\hat{\Gamma}_{1}^{(0)}$ by approximately ten percents. With the inclusion of QCD corrections, $\hat{\Gamma}_{1}^{(1)}$ significantly enhances the $\hat{\Gamma}_{1}^{(0)}$ for the $b\bar{b}[^3P_{0,1}^{1}]$ process; however, the enhancement is milder for $b\bar{b}[^3P_2^{1}]$. In terms of QED processes, $\hat{\Gamma}_{3}^{(1)}$ slightly diminishes the LO SDC $\hat{\Gamma}_{3}^{(0)}$ for $b\bar{b}[^3P_{0,1}^{1}]$, while it substantially reduces that for $b\bar{b}[^3P_2^{1}]$. Consequently, due to the combined influence of enhancement and reduction effects, the QCD corrections ($\hat{\Gamma}_{2}^{(1)}$) have a substantial impact on $\hat{\Gamma}_{2}^{(0)}$ corresponding to $b\bar{b}[^3P_{0,1}^{1}]$, whereas the effect is more moderate for $b\bar{b}[^3P_2^{1}]$.
\end{itemize}

The ratio $\hat\Gamma^{\textrm{NLO}}_{\textrm{total}}/\hat\Gamma^{\textrm{LO}}_{\textrm{total}}$ in Tables \ref{tab: 1} and \ref{tab: 2} suggests that QCD corrections can substantially enhance the LO results, underscoring the importance of our newly-calculated higher order terms in $\alpha_s$.

\begin{figure*}[!h]
\begin{center}
\hspace{0cm}\includegraphics[width=0.45\textwidth]{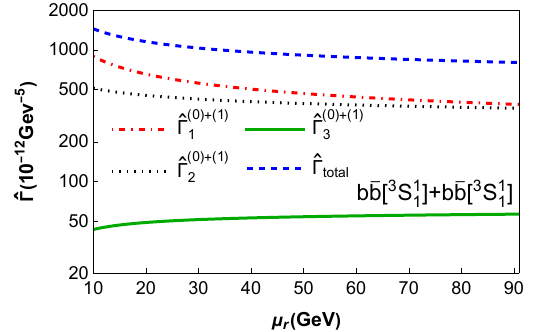}
\hspace{0cm}\includegraphics[width=0.45\textwidth]{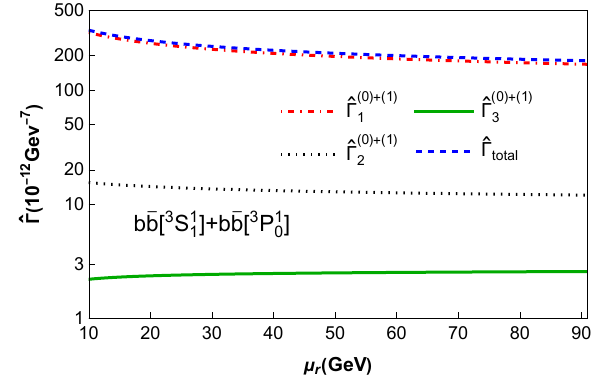}
\hspace{0cm}\includegraphics[width=0.45\textwidth]{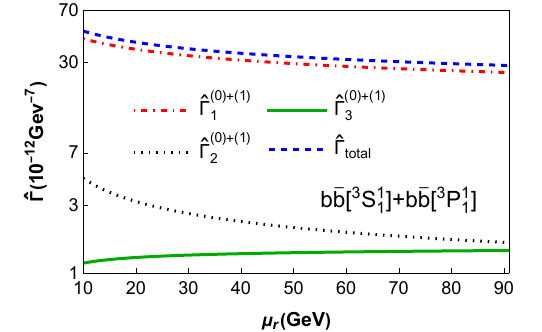}
\hspace{0cm}\includegraphics[width=0.45\textwidth]{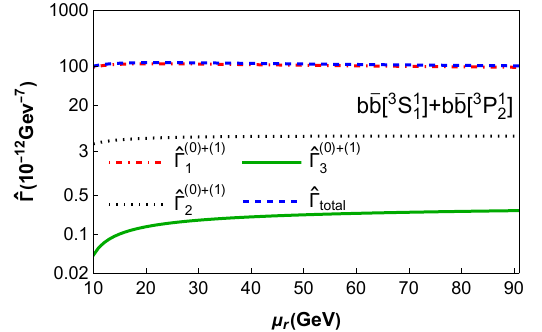}
\caption{\label{fig:SDC}
SDCs corresponding to $Z \to b\bar{b}[^3S_1^{1}]+b\bar{b}[^3S_1^{1},^3P_J^{1}]$ with respect to the renormalization scale $\mu_r$. The $b$-quark mass is fixed at $m_b=4.7$ GeV. $\hat{\Gamma}_{\textrm{total}}$ denotes the collected sum of the contributions of $\hat{\Gamma}_{1,2,3}^{(0,1)}$.}
\end{center}
\end{figure*}

To illustrate the relative importance of the QCD and QED contributions, along with their interference effects, we plot $\hat{\Gamma}_{0,1,2}$ as a function of the renormalization scale in Fig. \ref{fig:SDC}.

\begin{table*}[htb]
\begin{center}
\caption{Decay widths of $Z \to \Upsilon(mS)+\Upsilon(nS)$ (in unit: $10^{-12}$GeV) with $m_b=4.7$ GeV and $\mu_r=m_Z/2$. The subscripts ``dir" and ``fd" denote direct-production processes and feed-down effects, respectively. All predictions are composed of contributions from $\Gamma_{1,2,3}^{(0,1)}$.}
\label{tab: 3}
\begin{tabular}{|c|ccc|c|c|ccc|}
\hline
 & $\Gamma_{\textrm{dir}}$ & $\Gamma_{\textrm{fd}}$ & $\Gamma_{\textrm{total}}$ & ${\mathcal{B}}_\textrm{theo}$ & $\mathcal{B}_{\textrm{exp}}$ \cite{CMS:2022fsq}  \\ \hline
$1S+1S$ & $110.1$ & $43.58$ & $153.7$ & $6.158\times10^{-11}$ & $<1.8\times10^{-6}$ \\ \hline
$1S+2S$ & $109.9$ & $29.16$ & $139.1$ & $5.575\times10^{-11}$ & \\
$1S+3S$ & $84.10$ & $17.78$ & $101.9$ & $4.083\times10^{-11}$ & \\
$2S+2S$ & $27.45$ & $7.329$ & $34.78$ & $1.394\times10^{-11}$ & $<3.9\times10^{-7}$\\
$2S+3S$ & $41.99$ & $5.000$ & $46.99$ & $1.883\times10^{-11}$ & \\
$3S+3S$ & $16.06$ & $0.838$ & $16.90$ & $6.773\times10^{-12}$ & \\
\hline
\end{tabular}
\end{center}
\end{table*}

In Table \ref{tab: 3}, we confront our predictions with the CMS measurements. The subscripts ``dir" and ``fd" signify direct-production processes and feed-down effects, respectively. The ratio $\Gamma_{\textrm{fd}}/\Gamma_{\textrm{dir}}$ reveals that the feed-down contributions from the transitions of excited bottomonium states are crucial for $Z$ boson decay into a pair of $\Upsilon$ mesons. For example, in the case of $\Upsilon(1S)+\Upsilon(1S)$, feed-down contributions constitute approximately $30\%$ of the total results. Comparisons with experimental data indicate that the NLO predictions derived from the NRQCD framework are markedly below the upper threshold set by CMS.

Finally, we examine the uncertainties in our predictions resulting from variations in the $b$-quark mass $m_b$ and the renormalization scale $\mu_r$.
\begin{eqnarray}
\mathcal{B}_{Z \to \Upsilon(1S)+\Upsilon(1S)}&=&(6.158^{+0.254+2.868}_{-0.236-0.767})\times10^{-11}, \nonumber \\
\mathcal{B}_{Z \to \Upsilon(1S)+\Upsilon(2S)}&=&(5.575^{+0.233+2.586}_{-0.216-0.695})\times10^{-11}, \nonumber \\
\mathcal{B}_{Z \to \Upsilon(1S)+\Upsilon(3S)}&=&(4.083^{+0.168+1.905}_{-0.156-0.509})\times10^{-11}, \nonumber \\
\mathcal{B}_{Z \to \Upsilon(2S)+\Upsilon(2S)}&=&(1.394^{+0.059+0.647}_{-0.054-0.174})\times10^{-11}, \nonumber \\
\mathcal{B}_{Z \to \Upsilon(2S)+\Upsilon(3S)}&=&(1.883^{+0.078+0.877}_{-0.073-0.235})\times10^{-11}, \nonumber \\
\mathcal{B}_{Z \to \Upsilon(3S)+\Upsilon(3S)}&=&(6.773^{+0.277+3.173}_{-0.258-0.844})\times10^{-12},\label{eq19}
\end{eqnarray}
where the uncertainties listed in the first column stem from fluctuations in the value of $m_b$ between $4.6$ to $4.8$ GeV, centered around 4.7 GeV, whereas those in the second column result from adjustments to $\mu_r$ ranging from $2m_b$ to $m_Z$ around $m_Z/2$. The above results suggest that a 0.1 GeV variation in $m_b$ around 4.7 GeV lead to a modest change in predictions; however, adjusting $\mu_r$ from $2m_b$ to $m_Z$, particularly around $m_Z/2$, significantly affects the predictions.

Note that, in $Z$ boson decay into double $\Upsilon$, besides the color-singlet processes of interest, color-octet (CO) processes are also possible. The primary CO contributions are derived from the process $Z \to b\bar{b}[^3S_1^{8}]+g^{*}$, followed by $g^{*} \to b\bar{b}[^3S_1^{8}]$. Given the ratio of $\frac{\langle {\mathcal{O}}^{H}(^3S_{1}^{8}) \rangle}{\langle {\mathcal{O}}^{H}(^3S_{1}^{1}) \rangle} \sim 10^{-3}$ \cite{Br3}, the branching ratio $\mathcal{B}_{Z \to \Upsilon(1S)+\Upsilon(1S)}$ due to CO contributions are estimated to be $10^{-14}$. Moreover, considering that experiments account for feed down from higher excited states, the semi-inclusive production of double $\Upsilon$ mesons in $Z$ boson decay, such as contributions from $b$(or $\bar{b}$) quark fragmentation into $\Upsilon$, should be considered. Utilizing the fragmentation function from Ref. \cite{z decay 4}, the branching ratio $\mathcal{B}_{Z \to \Upsilon(1S)+\Upsilon(1S)+X}$ is predicted to be $\sim 10^{-11}$, which is comparable with the results in Eq. (\ref{eq19}).
\section{Summary}\label{sum}

In this paper, we carry out a comprehensive study of $Z \to \Upsilon(mS)+\Upsilon(nS)$ utilizing the NRQCD framework at the QCD NLO accuracy. The LO results indicate that the contributions from QED diagrams significantly enhance those from the QCD processes. The QCD corrections substantially amplify the QCD results, while concurrently reducing the QED results. Additionally, we discover that the transitions involving higher excited states of bottomonium mesons exert an indispensable influence. Taking into account both the QCD and QED contributions, we estimate the branching ratio for $\mathcal{B}_{Z \to \Upsilon(mS)+\Upsilon(nS)}$ to be on the order of $10^{-11}$. This value is significantly lower than the upper limit established by CMS and aligns  with experimental observations.

\appendix

\section{}

In this section, we present the expressions for the NLO coefficients $a_{i}$, $b_{i}$, and $c_{i}$ in Eq. (\ref{eq17}) which are formulated as a superposition of the MIs $\mathcal{I}$. The coefficients related to the $Z \to b\bar{b}[^3S_1^{1}]+b\bar{b}[^3S_1^{1}]$ process can be obtained by substituting $m_c$ in Eqs. (A1)-(A3) of Ref. \cite{z decay 40} with $m_b$. In the following, we only provide the expressions of $a_{i}$, $b_{i}$, and $c_{i}$ pertaining to the $Z \to b\bar{b}[^3S_1^{1}]+b\bar{b}[^3P_J^{1}]$ process.

\subsection{Master Integrals}
To start, we introduce the following definitions and showcase the finite ($\epsilon^0$-order) terms of the MIs that are involved ($r\equiv\frac{m^2_Z}{4m_b^2}$):

\begin{eqnarray}
&&a=\sqrt{r},~b=\sqrt{r -1},~c=\sqrt{r -4},~d=2 r+1, \nonumber \\ &&f=r+ac-4,~g=r-ac-4,~h=r g+2ac, \nonumber\\
&&j=r f-2ac,~j_1=(r-4) a b,~j_2=(2-r)b c.
\end{eqnarray}
There is only one 1-point scalar integral,
\begin{eqnarray}
\mathcal I_{1}&=&\frac{\lambda}{\mu_r^{4-D}}\int{\frac{d^D k}{k^2-m_b^2}}=m_b^2 \left[1-2\ln(m_b)\right], \label{kin}
\end{eqnarray}
where $k$ denotes the loop momentum, $\lambda=\frac{\mu_r^{4-D}}{i\pi^{\frac{D}{2}}\gamma_\Gamma}$ with $\gamma_\Gamma = \frac{\Gamma^2 (1-\epsilon)\Gamma (1+\epsilon)}{\Gamma (1-2\epsilon)}$.\\
There are five 2-point scalar integrals,
\begin{eqnarray}
\mathcal I^{(1)}_2&=&\frac{\lambda}{\mu_r^{4-D}}\int{\frac{d^D k}{k^2[(k+\frac{2p_1+p_2}{2})^2-m_b^2]}}=2-2\ln{(m_b)}-\frac{2r}{d} \left[\ln{(2r)}-i\pi\right], \label{MI1}
\end{eqnarray}
\begin{eqnarray}
\mathcal I^{(2)}_2&=&\frac{\lambda}{\mu_r^{4-D}}\int{\frac{d^D k}{k^2(k-\frac{p_1+p_2}{2})^2}}=2-\ln{({m_b}^2 r)}+i\pi, \label{MI2}
\end{eqnarray}
\begin{eqnarray}
\mathcal I^{(3)}_2&=&\frac{\lambda}{\mu_r^{4-D}}\int \frac{d^D k}{(k^2-m_b^2)[(k+\frac{p_1+p_2}{2})^2-m_b^2]}=\frac{c}{a}\left[\ln{ \left(\frac{4r}{(f+4)^2}\right)}+i\pi\right]+2\left[1-\ln{(m_b)}\right],\nonumber \\ \label{MI3}
\end{eqnarray}
\begin{eqnarray}
\mathcal I^{(4)}_2&=&\frac{\lambda}{\mu_r^{4-D}}\int\frac{d^D k}{(k^2-m_b^2)[(k+p_1+p_2)^2-m_b^2]} =\frac{b}{a}\left[\ln{(-2ab+d-2)}+i\pi\right]+2\left[1-\ln{(m_b)}\right],\nonumber \\ \label{MI4}
\end{eqnarray}
\begin{eqnarray}
\mathcal I^{(5)}_2&=&\frac{\lambda}{\mu^{4-D}}\int\frac{d^D k}{k^2(k-p_1)^2} = -\ln{\left(4m_b^2\right)}+i\pi+2,\label{MI5}
\end{eqnarray}
There are seven 3-point scalar integrals,
\begin{eqnarray}
&&\mathcal I^{(1)}_3\nonumber \\
&&=\frac{\lambda}{\mu_r^{4-D}}\int \frac{d^D k}{k^2[(k+\frac{p_2}{2})^2-m_b^2][(k+\frac{2p_1+p_2}{2})^2-m_b^2]} \nonumber\\
&&=\frac{1}{2 a c m_b^2}\left\{{\ln(2)\ln\left(\frac{(3(g+4)+h)^2}{d(g+h+4)^2}\right)+\left[\ln(r)-i\pi\right]\ln\left(\frac{d}{(g+3)^2}\right)+4 \textrm{Li}_2 \left(\frac{c}{a}\right)+\textrm{Li}_2 \left(-\frac{2(ac+j)}{d}\right)}\right.\nonumber\\
&&\left.{-\textrm{Li}_2 \left(\frac{c^2}{ac-h}\right)-\textrm{Li}_2 \left(-\frac{c+ag}{a}\right)-2\textrm{Li}_2 \left(-\frac{c}{a}\right)}\right\}, \label{MI6}
\end{eqnarray}
\begin{eqnarray}
&&\mathcal I^{(2)}_3\nonumber\\
&&=\frac{\lambda}{\mu_r^{4-D}}\int \frac{d^D k}{k^2[(k-\frac{p_2}{2})^2-m_b^2](k+\frac{p_1+p_2}{2})^2}\nonumber\\
&&=\frac{1}{ a c m_b^2}\left\lbrace{\ln(2)\ln\left(\frac{d(f+4)^2}{(3r-ac)^2}\right)+\left[\ln(r)-i\pi)\right]\ln\left(\frac{4dr}{(3r-ac)^2}\right)+\textrm{Li}_2\left(-\frac{2(ac+j)}{d}\right)+2\textrm{Li}_2 \left(\frac{c}{a}\right)}\right. \nonumber \\
&&\left.{+\textrm{Li}_2 \left(-\frac{g}{2}\right)-\textrm{Li}_2 \left(-\frac{f}{2}\right)-\textrm{Li}_2 \left(\frac{c^2}{ac-h}\right)-\textrm{Li}_2  \left(-\frac{c+ag}{a}\right)}\right\rbrace, \label{MI7}
\end{eqnarray}
\begin{eqnarray}
\mathcal I^{(3)}_3&=&\frac{\lambda}{\mu_r^{4-D}}\int \frac{d^D k}{k^2[(k+\frac{p_1}{2})^2-m_b^2](k+\frac{p_1+p_2}{2})^2} \nonumber\\
&=&\frac{1}{ a c m_b^2}\left\lbrace \left[\ln(r)-i\pi\right]\ln\left(\frac{(f+4)^2}{4r}\right)+2\textrm{Li}_2 \left(\frac{c}{a}\right)-2\textrm{Li}_2 \left(-\frac{c}{a}\right)-\textrm{Li}_2 \left(-\frac{g}{2}\right)+\textrm{Li}_2 \left(-\frac{f}{2}\right)\right\rbrace, \nonumber \\  \label{MI8}
\end{eqnarray}
\begin{eqnarray}
&&\mathcal I^{(4)}_3\nonumber\\
&&=\frac{\lambda}{\mu_r^{4-D}}\int \frac{d^D k}{k^2[(k+\frac{2p_1+p_2}{2})^2-m_b^2](k+\frac{p_1+p_2}{2})^2} \nonumber\\
&&= \frac{1}{ a c m_b^2}\left\lbrace{\ln(2)\ln\left(\frac{d}{(g+3)^2}\right)+\left[\ln(r)-i\pi)\right]\ln\left(\frac{g+6}{2g+6}\right)+\textrm{Li}_2 \left(-\frac{g}{2}\right)-\textrm{Li}_2 \left(-\frac{f}{2}\right)-2\textrm{Li}_2 \left(-\frac{c}{a}\right)}\right. \nonumber \\
&&\left.{-\textrm{Li}_2 \left(\frac{2ac-2h}{d}\right)+\textrm{Li}_2 \left(\frac{h-ac}{dr}\right)+\textrm{Li}_2 \left(\frac{c}{a}-f\right)}\right\rbrace, \label{MI9}
\end{eqnarray}
\begin{eqnarray}
&&\mathcal I^{(5)}_3\nonumber\\
&&=\frac{\lambda}{\mu_r^{4-D}}\int \frac{d^D k}{k^2[(k-\frac{p_1}{2})^2-m_b^2][(k+\frac{p_1+2p_2}{2})^2-m_b^2]}\nonumber\\
&& = \frac{1}{2 a c m_b^2}\left\{{\left[\ln\left(\frac{(r-ab)^2}{r}\right)+i\pi\right]\ln\left(\frac{br(2b-3c)+j-j_1}{br(2b+3c)+j+j_1}\right)+\left[\ln(2r)-i\pi\right]\ln\left(\frac{(g+3)^2}{d}\right)}\right. \nonumber \\
&&\left.{{-\textrm{Li}_2 \left(-\frac{2(ac+j)}{d}\right)-\textrm{Li}_2 \left(\frac{c^2 -h+j_1+j_2}{r}\right)-\textrm{Li}_2 \left(-\frac{-c^2 +h+j_1+j_2}{r}\right)+\textrm{Li}_2 \left(\frac{c^2 -j+j_1-j_2}{r}\right)}}\right.\nonumber \\
&&\left.{-2\textrm{Li}_2 \left(-\frac{c}{a}\right)+\textrm{Li}_2 \left(\frac{c^2 -j-j_1+j_2}{r}\right)+{\textrm{Li}_2 \left(\frac{c^2}{ac-h}\right)+\textrm{Li}_2 \left(-\frac{c+ag}{a}\right)}}\right\}, \nonumber \\ \label{MI10}
\end{eqnarray}
\begin{eqnarray}
&&\mathcal I^{(6)}_3\nonumber\\
&&=\frac{\lambda}{\mu_r^{4-D}}\int \frac{d^D k}{k^2[(k+\frac{p_2}{2})^2-m_b^2][(k+\frac{p_1+2p_2}{2})^2-m_b^2]} \nonumber\\
&&= \frac{1}{ a c m_b^2}\left\{{\ln(2)\ln \left( \frac{(f+4)^4\left(3(2r-ac)+h\right)}{4(g+6)\left(4(r-ac)+h\right)}\right)+\ln(r)\ln\left(\frac{dr}{(d+f+3)^2}\right)-i\pi\ln\left(\frac{16d}{(g+6)^2 r}\right)}\right.\nonumber \\
&&\left.{+\textrm{Li}_2 \left(-\frac{g}{8}\right)+\textrm{Li}_2 \left(-\frac{4ac+j}{2d}\right)-\textrm{Li}_2 \left(\frac{4ac-h}{4r}\right)+2\textrm{Li}_2 \left(\frac{c}{2a}\right)+\textrm{Li}_2 \left(\frac{f}{2r}\right)-\textrm{Li}_2 \left(-\frac{f}{8}\right)-\textrm{Li}_2 \left(\frac{g}{2r}\right)}\right.\nonumber \\
&&\left.{-\textrm{Li}_2 \left(\frac{c^2}{4ac-h}\right)}\right\}, \label{MI11}
\end{eqnarray}
\begin{eqnarray}
&&\mathcal I^{(7)}_3=\frac{\lambda}{\mu^{4-D}}\int \frac{d^D k}{k^2[(k+\frac{p_1}{2})^2-m_b^2](k-p2)^2} \nonumber\\
&&= \frac{1}{ 2 a c m_b^2}\left\{{\ln(2)\ln \left( \frac{64d}{(f+4)^4(g+6)^2}\right)+\ln(r)\ln\left(\frac{d(f+4)^4}{[(g+6)r]^2}\right)+i\pi\ln\left(\frac{(g+6)^2}{4d }\right)+\textrm{Li}_2 \left(-\frac{h}{2}\right)}\right.\nonumber \\
&&\left.{+2\textrm{Li}_2 \left(\frac{ac}{2}\right)-\textrm{Li}_2 \left(\frac{jr-2ac}{4d}\right)+\textrm{Li}_2 \left(\frac{2acr-jr^2}{2d}\right)-\textrm{Li}_2 \left(\frac{2ac-fr}{2}\right)-\textrm{Li}_2 \left(-\frac{2ac+hr}{4}\right)}\right\}.\nonumber \\
\label{MI12}
\end{eqnarray}

\subsection{NLO coefficients}
\subsubsection{$b\bar{b}[^3S_1^{1}]+b\bar{b}[^3P_J^{0}]$}
\begin{eqnarray}
a_1&=&-\frac{1}{3}{\mathcal I^{(2)}_2}-\frac{2 (r^4-25 r^3-146 r^2+176 r+72)}{9 (r^4+182 r^3-428 r^2+152 r+144)}-\frac{2}{3}\ln(2m_b), \label{A1}
\end{eqnarray}
\begin{eqnarray}
b_1&=&\frac{2(r^4+176 r^3-716 r^2+440 r+288)}{3 m_{b}^{2} r (r^4+182 r^3-428 r^2+152 r+144)}{\mathcal I_1}-\frac{r^5+184 r^4-76 r^3-1280 r^2+1024 r+576}{3 r (r^4+182 r^3-428 r^2+152 r+144)}{\mathcal I^{(3)}_2}\nonumber \\
&&-\frac{2 (r^5-13 r^4+814 r^3-2404 r^2+1176 r+864)}{9 r (r^4+182 r^3-428 r^2+152 r+144)}-\frac{2\ln(2m_b)}{3}, \label{A2}
\end{eqnarray}
\begin{eqnarray}
c_1&=&\frac{120r^7+19848r^6-103058r^5+54543r^4 +96834r^3 -16648r^2 -32352r-5760}{9m_b^2 r(r-4) (2r+1)(r^4+182r^3-428r^2+152r+144)}\mathcal I_{1}\nonumber \\
&&+\frac{5796 r^6-14278 r^5-21720 r^4+16167 r^3+32542 r^2+14136 r+2016}{9 r (2 r+1) (r^4+182 r^3-428 r^2+152 r+144)}\mathcal I^{(1)}_2\nonumber \\
&&+\frac{-1407 r^4+16354 r^3-25036 r^2+2136 r+6264}{9 (r^4+182 r^3-428 r^2+152 r+144)}\mathcal I^{(2)}_2\nonumber \\
&&+\frac{7 r^5+1408 r^4-8952 r^3+27008 r^2-20608 r-4992}{18 (r-4) (r^4+182 r^3-428 r^2+152 r+144)}\mathcal I^{(3)}_2\nonumber \\
&&-\frac{1469 r^4 + 2814 r^3 - 16728 r^2 + 11440 r + 4704}{9 (r^4+182 r^3-428 r^2+152 r+144)}\mathcal I^{(4)}_2 \nonumber \\
&&+\frac{2m_b^2 r (r^4 + 60 r^3 - 684 r^2 + 640 r + 240)}{3 (r^4+182 r^3-428 r^2+152 r+144)}\mathcal I^{(1)}_3\nonumber \\
&&+\frac{m_b^2 (-21 r^5 + 208 r^4 - 2137 r^3 + 2010 r^2 + 2444 r - 624)}{12 (r^4+182 r^3-428 r^2+152 r+144)}\mathcal I^{(2)}_3\nonumber \\
&&-\frac{m_b^2 (r^5 - 658 r^4 + 708 r^3 + 924 r^2 - 816 r - 288)}{3 (r^4+182 r^3-428 r^2+152 r+144)}\mathcal I^{(3)}_3\nonumber \\
&&+\frac{m_b^2 (r^5 - 480 r^4 - 3239 r^3 + 6838 r^2 - 1868 r - 528)}{12 (r^4+182 r^3-428 r^2+152 r+144)}\mathcal I^{(4)}_3\nonumber \\
&&-\frac{2m_b^2 (4 r^6 + 1208 r^5 - 6399 r^4 + 10840 r^3 - 6148 r^2 - 720 r +1344)}{3 (r-2) (r^4+182 r^3-428 r^2+152 r+144)}\mathcal I^{(5)}_3\nonumber \\
&&+\frac{2m_b^2 (3 r^5 + 96 r^4 - 406 r^3 + 860 r^2 - 528 r - 288)}{3 (r^4+182 r^3-428 r^2+152 r+144)}\mathcal I^{(6)}_3 \nonumber \\
&&-\frac{24m_b^2 r^2 (5 r^2 - 2 r - 6)}{(r-2) (r^4+182 r^3-428 r^2+152 r+144)}\mathcal I^{(7)}_3\nonumber \\
&&+\frac{(19 r^5+3512 r^4-7204 r^3+200 r^2+3968 r+768)}{r^5+182 r^4-428 r^3+152 r^2+144 r}\ln(m_b)\nonumber \\
&&+\frac{-25 r^6-12935 r^5+80893 r^4-118558 r^3-26064 r^2+85824 r+21888}{9 (r-4) r (r^4+182 r^3-428 r^2+152 r+144)}+11\ln(2), \nonumber \\
\label{A3}
\end{eqnarray}

\begin{eqnarray}
a_2&=&-\frac{1}{6}{\mathcal I^{(2)}_2}+\frac{7 r^4+38 r^3+16 r^2-424 r-144}{72 (7 r^4-7 r^3-59 r^2+56 r+36)}-\frac{\ln(2m_b)}{3},\label{A4}
\end{eqnarray}
\begin{eqnarray}
b_2&=&\frac{ 14 r^4 - 35 r^3 - 136 r^2 + 292 r + 144}{6 m_{b}^{2} r (7 r^4-7 r^3-59 r^2+56 r+36)}{\mathcal I_1}-\frac{ 7 r^5 + 7 r^4 - 94 r^3 - 80 r^2 + 328 r + 144}{6 r (7 r^4-7 r^3-59 r^2+56 r+36)}{\mathcal I^{(3)}_2}\nonumber \\
&&+\frac{7 r^5-256 r^4+328 r^3+2288 r^2-3216 r-1728}{72 r (7 r^4-7 r^3-59 r^2+56 r+36)}-\frac{\ln(2m_b)}{3},\label{A5}
\end{eqnarray}
\begin{eqnarray}
&&c_2=\nonumber\\
&&\frac{21672 r^8-69156 r^7-219454 r^6+398155 r^5+899784 r^4-46124 r^3-616672 r^2-309120 r-43776}{144 (r-4) r (2 r m_b+m_b)^2 (7 r^4-7 r^3-59 r^2+56 r+36)}\mathcal I_{1}\nonumber \\
&&+\frac{3840 r^8+34740 r^7-143528 r^6-28915 r^5+327531 r^4+350456 r^3+158500 r^2+33648 r+2880}{144 r (2 r+1)^2 (7 r^4-7 r^3-59 r^2+56 r+36)}\mathcal I^{(1)}_2\nonumber \\
&&+\frac{-240 r^5+714 r^4+13853 r^3-44450 r^2+13812 r+13176}{72 (7 r^4-7 r^3-59 r^2+56 r+36)}\mathcal I^{(2)}_2\nonumber \\
&&+\frac{56 r^5-1915 r^4+13254 r^3-30968 r^2+15472 r+8736}{36 (r-4) (7 r^4-7 r^3-59 r^2+56 r+36)}\mathcal I^{(3)}_2\nonumber \\
&&-\frac{ 120 r^5+1991 r^4-4392 r^3-1320 r^2+6016 r+2544}{36 (7 r^4-7 r^3-59 r^2+56 r+36)}\mathcal I^{(4)}_2\nonumber\\
&&+\frac{r (2 r^4 - 207 r^3 + 213 r^2 + 242 r + 12) m_b^2}{3 (7 r^4-7 r^3-59 r^2+56 r+36)}\mathcal I^{(1)}_3\nonumber \\
&&+\frac{(21 r^5 - 7333 r^4 + 22129 r^3 - 13620 r^2 - 140 r + 4368) m_b^2}{96 (7 r^4-7 r^3-59 r^2+56 r+36)}\mathcal I^{(2)}_3\nonumber \\
&&+\frac{ (100 r^5 + 965 r^4 - 5556 r^3 + 7692 r^2 - 1680 r - 2016) m_b^2}{24 (7 r^4-7 r^3-59 r^2+56 r+36)}\mathcal I^{(3)}_3\nonumber \\
&&-\frac{(13 r^5 - 3993 r^4 + 1633 r^3 + 12124 r^2 - 11084 r - 3696) m_b^2}{96 (7 r^4-7 r^3-59 r^2+56 r+36)}\mathcal I^{(4)}_3\nonumber \\
&&-\frac{(628 r^6 - 1309 r^5 - 4890 r^4 + 14188 r^3 - 5944 r^2 - 4704 r +1920) m_b^2}{24 (r-2) (7 r^4-7 r^3-59 r^2+56 r+36)}\mathcal I^{(5)}_3\nonumber \\
&&+\frac{ (15 r^5 - 171 r^4 + 2044 r^3 - 6236 r^2 + 3552 r + 2016) m_b^2}{12 (7 r^4-7 r^3-59 r^2+56 r+36)}\mathcal I^{(6)}_3\nonumber \\
&&-\frac{3r^2 (3 r^3 + 2 r^2 - 20 r - 24)  m_b^2}{4 (r-2) (7 r^4-7 r^3-59 r^2+56 r+36)}\mathcal I^{(7)}_3\nonumber \\
&&+\frac{759 r^5-708 r^4-6736 r^3+3312 r^2+7312 r+1536}{8 r (7 r^4-7 r^3-59 r^2+56 r+36)}\ln(m_b)\nonumber \\
&&+\frac{-16562 r^7+78949 r^6+154533 r^5-887080 r^4-10228 r^3+1040208 r^2+535104 r+66816}{144 (r-4) r (2 r+1) (7 r^4-7 r^3-59 r^2+56 r+36)}\nonumber \\
&&+\frac{11}{2}\ln(2),\label{A6}
\end{eqnarray}

\begin{eqnarray}
a_3&=&b_3=0,\label{A7}
\end{eqnarray}
\begin{eqnarray}
&&c_3\nonumber\\
&&=\frac{\mathcal I_{1}}{9 (r-4) r (2 r m_b+m_b)^2 (9 r^5+2 r^4-152 r^3-16 r^2+592 r+288)}\nonumber \\
&&\times(1512 r^9+3216 r^8-31716 r^7+29233 r^6+208412 r^5+192816 r^4-374944 r^3\nonumber \\
&&-536912 r^2-231744 r-32256)\nonumber \\
&&+\frac{ 4(1446 r^8 - 1323 r^7 - 10786 r^6 + 27874 r^5 + 53997 r^4 +
26482 r^3 + 2900 r^2 - 1320 r - 288)}{9 r (2 r+1)^2 (9 r^5+2 r^4-152 r^3-16 r^2+592 r+288)}\mathcal I^{(1)}_2\nonumber \\
&&-\frac{ 8(60 r^5 - 429 r^4 - 70 r^3 + 3868 r^2 - 3000 r - 1728)}{9 (9 r^5+2 r^4-152 r^3-16 r^2+592 r+288)}\mathcal I^{(2)}_2\nonumber \\
&&-\frac{64(17 r^5-103 r^4+51 r^3+850 r^2-1640 r-624)}{9 (r-4) (9 r^5+2 r^4-152 r^3-16 r^2+592 r+288)}\mathcal I^{(3)}_2\nonumber \\
&&-\frac{4 (255 r^5-8 r^4-1272 r^3+1392 r^2-784 r+192)}{9 (9 r^5+2 r^4-152 r^3-16 r^2+592 r+288)}\mathcal I^{(4)}_2\nonumber\\
&&-\frac{8r (55 r^4 - 24 r^3 - 480 r^2 - 56 r + 96) m_b^2}{3 (9 r^5+2 r^4-152 r^3-16 r^2+592 r+288)}\mathcal I^{(1)}_3\nonumber \\
&&-\frac{ (273 r^5 - 484 r^4 - 2138 r^3 + 3408 r^2 - 56 r - 1248) m_b^2}{3 (9 r^5+2 r^4-152 r^3-16 r^2+592 r+288)}\mathcal I^{(2)}_3\nonumber \\
&&+\frac{ 2(155 r^5 - 530 r^4 - 528 r^3 + 4728 r^2 - 3792 r - 2304) m_b^2}{3 (9 r^5+2 r^4-152 r^3-16 r^2+592 r+288)}\mathcal I^{(3)}_3\nonumber \\
&&+\frac{4(86 r^5 + 117 r^4 - 805 r^3 - 892 r^2 + 1916 r + 528) m_b^2}{3 (9 r^5+2 r^4-152 r^3-16 r^2+592 r+288)}\mathcal I^{(4)}_3\nonumber \\
&&-\frac{4(18 r^6 + 85 r^5 - 32 r^4 - 772 r^3 + 272 r^2 + 1344 r + 192) m_b^2}{24 (r-2) (3 (9 r^5+2 r^4-152 r^3-16 r^2+592 r+288)}\mathcal I^{(5)}_3\nonumber \\
&&-\frac{ 32 (3 r^5 - 42 r^4 + 20 r^3 + 320 r^2 - 672 r - 288) m_b^2}{3 (9 r^5+2 r^4-152 r^3-16 r^2+592 r+288)}\mathcal I^{(6)}_3\nonumber \\
&&+\frac{2 (36 r^6 - 7 r^5 - 668 r^4 - 168 r^3 + 2608 r^2 + 3344 r + 768) }{r (9 r^5+2 r^4-152 r^3-16 r^2+592 r+288)}\ln(m_b)\nonumber \\
&&+\frac{-1404 r^8+5174 r^7+34583 r^6-121380 r^5-207200 r^4+497824 r^3+727440 r^2+265536 r+23040}{9 (r-4) r (2 r+1) (9 r^5+2 r^4-152 r^3-16 r^2+592 r+288)}.\label{A8}\nonumber\\
\end{eqnarray}

\subsubsection{$b\bar{b}[^3S_1^{1}]+b\bar{b}[^3P_1^{1}]$}

\begin{eqnarray}
a_1&=&-\frac{1}{3}{\mathcal I^{(2)}_2}-\frac{2 r^3-15 r^2+16 r+36}{18 (2 r^3-12 r^2+13 r+18)}-\frac{2\ln(2m_b)}{3},\label{A9}
\end{eqnarray}

\begin{eqnarray}
b_1&=&\frac{ 4(r^3-9 r^2+14 r+18)}{3 m_{b}^{2} r (2 r^3 - 12 r^2 + 13 r + 18)}{\mathcal I_1}-\frac{  2 r^4 - 8 r^3 - 23 r^2 + 74 r + 72}{3 r (2 r^3-12 r^2+13 r+18)}{\mathcal I^{(3)}_2}\nonumber \\
&&-\frac{2 r^4+45 r^3-290 r^2+264 r+432}{18 r (2 r^3-12 r^2+13 r+18)}-\frac{2\ln(2m_b)}{3},\label{A10}
\end{eqnarray}

\begin{eqnarray}
c_1&=&-\frac{ (-784 r^7 + 4412 r^6 - 3080 r^5 - 9905 r^4 + 5302 r^3 + 13190 r^2 +
  6456 r + 960)}{12 (r-4) r (2 r^3-12 r^2+13 r+18) (2 m_b r+m_b)^2}\mathcal I_{1}\nonumber \\
&&+\frac{64 r^7-2592 r^6+396 r^5+12116 r^4+16867 r^3+10206 r^2+2930 r+336}{12 r (2 r+1)^2 (2 r^3-12 r^2+13 r+18)}\mathcal I^{(1)}_2\nonumber \\
&&+\frac{-4 r^4+323 r^3-877 r^2-134 r+522}{12 r^3-72 r^2+78 r+108}\mathcal I^{(2)}_2\nonumber \\
&&+\frac{19 r^4-291 r^3+1227 r^2-1180 r-624}{18 (r-4) (2 r^3-12 r^2+13 r+18)}\mathcal I^{(3)}_2\nonumber \\
&&+\frac{-12 r^4+74 r^3+473 r^2-2020 r-1176}{18 (2 r^3-12 r^2+13 r+18)}\mathcal I^{(4)}_2\nonumber\\
&&+\frac{2m_b^2 r (4 r^3 - 23 r^2 + 41 r + 30)}{6 r^3-36 r^2+39 r+54}\mathcal I^{(1)}_3\nonumber \\
&&-\frac{m_b^2 (280 r^4 + 450 r^3 - 449 r^2 - 1040 r + 312)}{48 (2 r^3-12 r^2+13 r+18)}\mathcal I^{(2)}_3\nonumber \\
&&+\frac{m_b^2 (34 r^4 - 135 r^3 - 129 r^2 + 276 r + 144)}{12 (2 r^3-12 r^2+13 r+18)}\mathcal I^{(3)}_3\nonumber \\
&&-\frac{ m_b^2 (408 r^4 - 1010 r^3 - 2523 r^2 + 512 r + 264)}{48 (2 r^3-12 r^2+13 r+18)}\mathcal I^{(4)}_3\nonumber \\
&&-\frac{m_b^2 (22 r^4 - 247 r^3 + 635 r^2 + 146 r - 336)}{6 (2 r^3-12 r^2+13 r+18)}\mathcal I^{(5)}_3\nonumber \\
&&-\frac{m_b^2 (10 r^4 - 25 r^3 - 162 r^2 + 168 r + 144)}{6 (2 r^3-12 r^2+13 r+18)}\mathcal I^{(6)}_3\nonumber \\
&&+\frac{40 r^4-223 r^3+171 r^2+444 r+96}{2 r^4-12 r^3+13 r^2+18 r}\ln(m_b)\nonumber \\
&&+\frac{-598r^6+6783r^5-19185r^4-5999r^3+40917r^2+29940r+5472}{18(r-4)r(2 r+1)(2 r^3-12 r^2+13 r+18)}+11\ln(2),\nonumber\\
\label{A11}
\end{eqnarray}

\begin{eqnarray}
a_2&=&-\frac{1}{6}{\mathcal I^{(2)}_2}+\frac{-5 r^3+24 r^2+68 r+72}{18 (19 r^3-24 r^2-88 r-72)}-\frac{\ln(2m_b)}{3},\label{A12}
\end{eqnarray}

\begin{eqnarray}
b_2&=&\frac{ r (r (19 r-72)-184)-144}{3 m_{b}^{2} r (r (r (19 r - 24) - 88) - 72)}{\mathcal I_1}-\frac{  19 r^4+14 r^3-232 r^2-440 r-288}{6 r (19 r^3-24 r^2-88 r-72)}{\mathcal I^{(3)}_2}\nonumber \\
&&+\frac{-5 r^4-234 r^3+212 r^2+960 r+864}{18 r (19 r^3-24 r^2-88 r-72)}-\frac{\ln(2m_b)}{3},\label{A13}
\end{eqnarray}

\begin{eqnarray}
&&c_2\nonumber\\
&&=\frac{116 r^8+2088 r^7-12033 r^6-6098 r^5+10386 r^4+72762 r^3+94260 r^2+45360 r+7296}{12 (r-4) r (19 r^3-24 r^2-88 r-72) (2 m_b r+m_b)^2}\mathcal I_{1}\nonumber \\
&&-\frac{128 r^8+332 r^7-3272 r^6+31989 r^5+76948 r^4+67272 r^3+28738 r^2+5900 r+480}{12 r (2 r+1)^2 (19 r^3-24 r^2-88 r-72)}\mathcal I^{(1)}_2\nonumber \\
&&+\frac{-16 r^5+113 r^4-838 r^3-5202 r^2+1820 r+4392}{12 (-19 r^3+24 r^2+88 r+72)}\mathcal I^{(2)}_2\nonumber \\
&&-\frac{9 r^5-607 r^4+4602 r^3-17004 r^2+15664 r+17472}{36 (r-4) (19 r^3-24 r^2-88 r-72)}\mathcal I^{(3)}_2\nonumber \\
&&+\frac{24 r^5+207 r^4-1346 r^3+6268 r^2+6448 r+2544}{18 (19 r^3-24 r^2-88 r-72)}\mathcal I^{(4)}_2\nonumber\\
&&-\frac{2m_b^2 r (3 r^4 - 62 r^3 + 121 r^2 + 131 r + 12)}{3 (19 r^3-24 r^2-88 r-72)}\mathcal I^{(1)}_3\nonumber \\
&&+\frac{m_b (492 r^5 + 2098 r^4 - 4215 r^3 + 2314 r^2 + 184 r - 4368)}{48 (19 r^3-24 r^2-88 r-72)}\mathcal I^{(2)}_3\nonumber \\
&&-\frac{m_b (18 r^5 - 119 r^4 - 1203 r^3 + 2922 r^2 - 480 r - 2016)}{12 (19 r^3-24 r^2-88 r-72)}\mathcal I^{(3)}_3\nonumber \\
&&-\frac{636 r^5+1578 r^4+6869 r^3-2382 r^2+5176 r+3696}{48 (19 r^3-24 r^2-88 r-72)}\mathcal I^{(4)}_3\nonumber \\
&&+\frac{m_b (54 r^5 - 458 r^4 + 1745 r^3 + 728 r^2 + 788 r - 480)}{6 (19 r^3-24 r^2-88 r-72)}\mathcal I^{(5)}_3\nonumber \\
&&-\frac{ m_b (6 r^5 + 41 r^4 + 460 r^3 - 1764 r^2 + 2208 r + 2016)}{6 (19 r^3-24 r^2-88 r-72)}\mathcal I^{(6)}_3\nonumber \\
&&+\frac{-539 r^4+442 r^3+2656 r^2+3240 r+768}{-38 r^4+48 r^3+176 r^2+144 r}\ln(m_b)\nonumber \\
&&-\frac{-468 r^7+20074 r^6-70839 r^5-100506 r^4+248630 r^3+519516 r^2+256080 r+33408}{36 (r-4) r (2 r+1) (19 r^3-24 r^2-88 r-72)}+\frac{11}{2}\ln(2),\nonumber\\\label{A14}
\end{eqnarray}

\begin{eqnarray}
a_3&=&b_3=0,\label{A15}
\end{eqnarray}

\begin{eqnarray}
&&c_3\nonumber\\
&&=-\frac{648 r^9-1628 r^8-200 r^7-5270 r^6-29735 r^5-58006 r^4-89032 r^3-79760 r^2-34176 r-5376}{3 (r-4) r (2 r m_b+m_b)^2 (9 r^5-9 r^4+34 r^3+228 r^2+248 r+144)}\mathcal I_{1}\nonumber \\
&&+\frac{2 (28 r^8 - 1626 r^7 + 1334 r^6 + 14055 r^5 + 19662 r^4 + 11099 r^3 +
2300 r^2 - 196 r - 96)}{3 r (2 r+1)^2 (9 r^5-9 r^4+34 r^3+228 r^2+248 r+144)}\mathcal I^{(1)}_2\nonumber \\
&&-\frac{4(4 r^5-93 r^4+276 r^3+140 r^2-616 r-576)}{3 (9 r^5-9 r^4+34 r^3+228 r^2+248 r+144)}\mathcal I^{(2)}_2\nonumber \\
&&+\frac{8 (-9 r^5 + 269 r^4 - 1230 r^3 + 1068 r^2 + 3952 r + 2496)}{9 (r-4) (9 r^5-9 r^4+34 r^3+228 r^2+248 r+144)}\mathcal I^{(3)}_2\nonumber \\
&&-\frac{4 (39 r^5 - 219 r^4 + 1102 r^3 + 1960 r^2 + 664 r + 96)}{9 (9 r^5-9 r^4+34 r^3+228 r^2+248 r+144)}\mathcal I^{(4)}_2\nonumber\\
&&-\frac{4 r (57 r^4 - 49 r^3 - 127 r^2 + 10 r + 96) m_b^2}{3 (9 r^5-9 r^4+34 r^3+228 r^2+248 r+144)}\mathcal I^{(1)}_3\nonumber \\
&&+\frac{2(6 r^5 + 71 r^4 - 174 r^3 - 226 r^2 + 68 r + 624) m_b^2}{3 (9 r^5-9 r^4+34 r^3+228 r^2+248 r+144)}\mathcal I^{(2)}_3\nonumber \\
&&+\frac{2(36 r^5 - 161 r^4 + 546 r^3 + 294 r^2 - 1164 r - 1152) m_b^2}{3 (9 r^5-9 r^4+34 r^3+228 r^2+248 r+144)}\mathcal I^{(3)}_3\nonumber \\
&&-\frac{4 (21 r^5 - 72 r^4 + 80 r^3 + 12 r^2 - 536 r - 264) m_b^2}{3 (9 r^5-9 r^4+34 r^3+228 r^2+248 r+144)}\mathcal I^{(4)}_3\nonumber \\
&&-\frac{4(18 r^6 - 45 r^5 + 271 r^4 + 476 r^3 + 188 r^2 + 464 r + 96) m_b^2}{3 (9 r^5-9 r^4+34 r^3+228 r^2+248 r+144)}\mathcal I^{(5)}_3\nonumber \\
&&-\frac{16 (6 r^5 - 11 r^4 + 50 r^3 - 144 r^2 - 480 r - 288) m_b^2}{3 (9 r^5-9 r^4+34 r^3+228 r^2+248 r+144)}\mathcal I^{(6)}_3\nonumber \\
&&+\frac{72 r^6-72 r^5+222 r^4+2200 r^3+3440 r^2+2928 r+768}{9 r^6-9 r^5+34 r^4+228 r^3+248 r^2+144 r}\ln(m_b)\nonumber \\
&&+\frac{-1944 r^8+10224 r^7-9718 r^6-62979 r^5+122574 r^4+343096 r^3+350880 r^2+131136 r+11520}{9 (r-4) r (2 r+1) (9 r^5-9 r^4+34 r^3+228 r^2+248 r+144)}.\nonumber\\
\label{A16}
\end{eqnarray}

\subsubsection{$b\bar{b}[^3S_1^{1}]+b\bar{b}[^3P_2^{1}]$}

\begin{eqnarray}
a_1&=&-\frac{1}{3}{\mathcal I^{(2)}_2}-\frac{2 (r^4+11 r^3-131 r^2+224 r+180)}{9 (r^4+20 r^3-188 r^2+308 r+360)}-\frac{2\ln(2m_b)}{3},\label{A17}
\end{eqnarray}

\begin{eqnarray}
b_1&=&\frac{ 2 (r^4 + 14 r^3 - 296 r^2 + 776 r + 720)}{3 m_{b}^{2}  r (r^4 + 20 r^3 - 188 r^2 + 308 r + 360)}{\mathcal I_1}-\frac{  r^5+22 r^4-160 r^3-284 r^2+1912 r+1440}{3 r (r^4+20 r^3-188 r^2+308 r+360)}{\mathcal I^{(3)}_2}\nonumber \\
&&-\frac{2 (r^5+23 r^4+73 r^3-1366 r^2+1968 r+2160)}{9 r (r^4+20 r^3-188 r^2+308 r+360)}-\frac{2\ln(2m_b)}{3},\label{A18}
\end{eqnarray}

\begin{eqnarray}
&&c_1\nonumber\\
&&=\frac{204 r^8+1812 r^7-57877 r^6+146122 r^5+217023 r^4-115334 r^3-239518 r^2-104568 r-14400}{9 (r-4) r (2 r m_b+m_b)^2 (r^4+20 r^3-188 r^2+308 r+360)}\mathcal I_{1}\nonumber \\
&&+\frac{1248 r^7+6628 r^6+127916 r^5+289545 r^4+290345 r^3+156466 r^2+43890 r+5040}{9 r (2 r + 1)^2 (r^4 + 20 r^3 - 188 r^2 + 308 r + 360}\mathcal I^{(1)}_2\nonumber \\
&&+\frac{-123 r^4+868 r^3-28744 r^2-2868 r+15660}{9 (r^4+20 r^3-188 r^2+308 r+360)}\mathcal I^{(2)}_2\nonumber \\
&&-\frac{101 r^5+146 r^4-2496 r^3-260 r^2+1936 r+12480}{18 (r-4) (r^4+20 r^3-188 r^2+308 r+360)}\mathcal I^{(3)}_2\nonumber \\
&&-\frac{113 r^4+798 r^3+5286 r^2+17944 r+11760}{9 (r^4+20 r^3-188 r^2+308 r+360)}\mathcal I^{(4)}_2\nonumber \\
&&+\frac{r (r^3 + 14 r^2 - 78 r - 72) }{r^4+20 r^3-188 r^2+308 r+360}\mathcal I^{(5)}_2\nonumber \\
&&+\frac{2r (r^4 + 24 r^3 - 306 r^2 + 340 r + 600) m_b^2}{3 (r^4+20 r^3-188 r^2+308 r+360)}\mathcal I^{(1)}_3\nonumber \\
&&-\frac{(21 r^5 + 1115 r^4 + 1003 r^3 - 3945 r^2 - 2324 r + 1560) m_b^2}{12 (r^4+20 r^3-188 r^2+308 r+360)}\mathcal I^{(2)}_3\nonumber \\
&&-\frac{(r^5 - 109 r^4 + 498 r^3 + 369 r^2 - 996 r - 720) m_b^2}{3 (r^4+20 r^3-188 r^2+308 r+360)}\mathcal I^{(3)}_3\nonumber \\
&&+\frac{ (r^5 + 1251 r^4 + 14599 r^3 + 9475 r^2 + 700 r - 1320) m_b^2}{12 (r^4+20 r^3-188 r^2+308 r+360)}\mathcal I^{(4)}_3\nonumber \\
&&-\frac{2 (4 r^6 + 209 r^5 - 1014 r^4 + 4477 r^3 - 5416 r^2 - 3468 r +
   3360) m_b^2}{3 (r-2) (r^4+20 r^3-188 r^2+308 r+360)}\mathcal I^{(5)}_3\nonumber \\
&&+\frac{2 (3 r^5 + 45 r^4 - 25 r^3 + 230 r^2 - 600 r - 720) m_b^2}{3 (r^4+20 r^3-188 r^2+308 r+360)}\mathcal I^{(6)}_3\nonumber \\
&&-\frac{72(r - 4) r (r + 1) m_b^2}{(r-2) (r^4+20 r^3-188 r^2+308 r+360)}\mathcal I^{(7)}_3\nonumber \\
&&+\frac{19 r^5+416 r^4-3664 r^3+4940 r^2+9296 r+1920}{r^5+20 r^4-188 r^3+308 r^2+360 r}\ln(m_b)\nonumber \\
&&+\frac{-104 r^7-3176 r^6+29871 r^5-103006 r^4+32774 r^3+441486 r^2+306648 r+54720}{9 (r-4) r (2 r+1) (r^4+20 r^3-188 r^2+308 r+360)}\nonumber \\
&&+11\ln(2).\label{A19}
\end{eqnarray}

\begin{eqnarray}
a_2&=&-\frac{1}{6}{\mathcal I^{(2)}_2}+\frac{-11 r^4+65 r^3+148 r^2-628 r-360}{18 (10 r^4-37 r^3-104 r^2+488 r+360)}-\frac{\ln(2m_b)}{3},\label{A20}
\end{eqnarray}

\begin{eqnarray}
b_2&=&\frac{ 10 r^4-79 r^3-68 r^2+1136 r+720}{3 m_{b}^{2} r (10 r^4 - 37 r^3 - 104 r^2 + 488 r + 360)}{\mathcal I_1}-\frac{  10 r^5-17 r^4-262 r^3+352 r^2+2632 r+1440}{6 r (10 r^4-37 r^3-104 r^2+488 r+360)}{\mathcal I^{(3)}_2}\nonumber \\
&&-\frac{11 r^5+121 r^4-730 r^3-1724 r^2+6096 r+4320}{18 r (10 r^4-37 r^3-104 r^2+488 r+360)}-\frac{\ln(2m_b)}{3},\label{A21}
\end{eqnarray}

\begin{eqnarray}
&&c_2\nonumber\\
&&=\frac{3096 r^8-24996 r^7-30164 r^6+323237 r^5+146433 r^4-521269 r^3-663026 r^2-317208 r-54720}{18 (r - 4) r (2 r m_b + m_b)^2 (10 r^4 - 37 r^3 - 104 r^2 + 488 r + 360}\mathcal I_{1}\nonumber \\
&&+\frac{1536 r^8+17268 r^7+111220 r^6+635537 r^5+1298688 r^4+1176860 r^3+526618 r^2+107484 r+7200}{36 r (2 r + 1)^2 (10 r^4 - 37 r^3 - 104 r^2 + 488 r + 360}\nonumber\\
&&\times\mathcal I^{(1)}_2-\frac{192 r^5+1041 r^4+12656 r^3+93670 r^2-18084 r-65880}{36 (10 r^4-37 r^3-104 r^2+488 r+360)}\mathcal I^{(2)}_2\nonumber \\
&&+\frac{-289 r^5-193 r^4+9762 r^3-36644 r^2+65680 r+87360}{36 (r-4) (10 r^4-37 r^3-104 r^2+488 r+360)}\mathcal I^{(3)}_2\nonumber \\
&&-\frac{48 r^5+401 r^4+2466 r^3+11496 r^2+23680 r+6360}{9 (10 r^4-37 r^3-104 r^2+488 r+360)}\mathcal I^{(4)}_2\nonumber \\
&&+\frac{ r (2 r^3 - 8 r^2 - 57 r - 36) }{10 r^4-37 r^3-104 r^2+488 r+360}\mathcal I^{(5)}_2\nonumber \\
&&+\frac{2r (4 r^4 - 189 r^3 + 48 r^2 + 247 r + 60) m_b^2}{3 (10 r^4-37 r^3-104 r^2+488 r+360)}\mathcal I^{(1)}_3\nonumber \\
&&-\frac{(822 r^5 + 4154 r^4 - 12011 r^3 + 14334 r^2 - 13424 r - 21840) m_b^2}{48 (10 r^4-37 r^3-104 r^2+488 r+360)}\mathcal I^{(2)}_3\nonumber \\
&&+\frac{ (56 r^5 + 427 r^4 - 4071 r^3 + 8370 r^2 + 384 r - 10080) m_b^2}{12 (10 r^4-37 r^3-104 r^2+488 r+360)}\mathcal I^{(3)}_3\nonumber \\
&&+\frac{ (1414 r^5 + 12714 r^4 + 50113 r^3 + 13030 r^2 - 4208 r + 18480) m_b^2}{48 (10 r^4-37 r^3-104 r^2+488 r+360)}\mathcal I^{(4)}_3\nonumber \\
&&-\frac{(340 r^6 - 796 r^5 + 3933 r^4 + 6358 r^3 - 28252 r^2 + 1224 r +
   4800) m_b^2}{6 (r-2) (10 r^4-37 r^3-104 r^2+488 r+360)}\mathcal I^{(5)}_3\nonumber \\
&&+\frac{(30 r^5 + 21 r^4 + 1760 r^3 - 3004 r^2 + 7680 r + 10080) m_b^2}{6 (10 r^4-37 r^3-104 r^2+488 r+360)}\mathcal I^{(6)}_3\nonumber \\
&&-\frac{18r (r^3 - 10 r - 8) m_b^2}{(r-2) (10 r^4-37 r^3-104 r^2+488 r+360)}\mathcal I^{(7)}_3\nonumber \\
&&+\frac{291 r^5-915 r^4-4006 r^3+13632 r^2+17032 r+3840}{20 r^5-74 r^4-208 r^3+976 r^2+720 r}\ln(m_b)\nonumber \\
&&+\frac{-7688 r^7+41692 r^6+62841 r^5-824998 r^4+799850 r^3+2723892 r^2+1386288 r+167040}{36 (r-4) r (2 r+1) (10 r^4-37 r^3-104 r^2+488 r+360)}\nonumber\\
&&+\frac{11}{2}\ln(2),
\label{A22}
\end{eqnarray}

\begin{eqnarray}
a_3&=&b_3=0,\label{A23}
\end{eqnarray}

\begin{eqnarray}
&&c_3\nonumber \\
&&=-\frac{1}{9 (r-4) r (2 r m_b+m_b)^2 (9 r^5+11 r^4-134 r^3+140 r^2+1336 r+720)}\nonumber \\
&&\times(-1512 r^9-1032 r^8+32016 r^7-11119 r^6-117845 r^5+317166 r^4+933424 r^3+943568 r^2\nonumber \\
&&+451200 r+80640)\mathcal I_{1}\nonumber \\
&&+\frac{2(444 r^8+4470 r^7+24088 r^6+165809 r^5+332502 r^4+265415 r^3+89104 r^2+6612 r-1440)}{9 r (2 r+1)^2 (9 r^5+11 r^4-134 r^3+140 r^2+1336 r+720)}\mathcal I^{(1)}_2\nonumber \\
&&-\frac{4 (48 r^5 - 105 r^4 + 952 r^3 + 7196 r^2 - 6360 r - 8640)}{9 (9 r^5+11 r^4-134 r^3+140 r^2+1336 r+720)}\mathcal I^{(2)}_2\nonumber \\
&&-\frac{8 (55 r^5 - 281 r^4 + 462 r^3 + 2516 r^2 - 15472 r - 12480)}{9 (r-4) (9 r^5+11 r^4-134 r^3+140 r^2+1336 r+720)}\mathcal I^{(3)}_2\nonumber \\
&&-\frac{4 (21 r^5 + 619 r^4 + 2706 r^3 + 10896 r^2 + 12008 r + 480)}{9 (9 r^5+11 r^4-134 r^3+140 r^2+1336 r+720)}\mathcal I^{(4)}_2\nonumber \\
&&-\frac{4r (110 r^4 + 249 r^3 - 339 r^2 + 386 r + 480) m_b^2}{3 (9 r^5+11 r^4-134 r^3+140 r^2+1336 r+720)}\mathcal I^{(1)}_3\nonumber \\
&&-\frac{2(57 r^5 - 367 r^4 - 608 r^3 + 3210 r^2 - 2852 r - 3120) m_b^2}{3 (9 r^5+11 r^4-134 r^3+140 r^2+1336 r+720)}\mathcal I^{(2)}_3\nonumber \\
&&+\frac{2 (83 r^5 - 125 r^4 - 336 r^3 + 4554 r^2 - 4044 r - 5760) m_b^2}{3 (9 r^5+11 r^4-134 r^3+140 r^2+1336 r+720)}\mathcal I^{(3)}_3\nonumber \\
&&+\frac{8(7 r^5 + 93 r^4 + 376 r^3 - 347 r^2 + 196 r + 660) m_b^2}{3 (9 r^5+11 r^4-134 r^3+140 r^2+1336 r+720)}\mathcal I^{(4)}_3\nonumber \\
&&-\frac{4 (18 r^6 + 103 r^5 + 247 r^4 + 2660 r^3 + 2348 r^2 - 720 r + 480) m_b^2}{3 (9 r^5+11 r^4-134 r^3+140 r^2+1336 r+720)}\mathcal I^{(5)}_3\nonumber \\
&&-\frac{16 (6 r^5 - 51 r^4 - 80 r^3 + 28 r^2 - 1920 r - 1440) m_b^2}{3 (9 r^5+11 r^4-134 r^3+140 r^2+1336 r+720)}\mathcal I^{(6)}_3\nonumber \\
&&+\frac{2 (36 r^6 + 101 r^5 - 731 r^4 - 60 r^3 + 7432 r^2 + 7736 r + 1920)}{r (9 r^5+11 r^4-134 r^3+140 r^2+1336 r+720)}\ln(m_b)\nonumber \\
&&+\frac{-2052 r^8+4274 r^7+57269 r^6-136869 r^5-528734 r^4+1033984 r^3+1897728 r^2+747072 r+57600}{9 (r-4) r (2 r+1) (9 r^5+11 r^4-134 r^3+140 r^2+1336 r+720)}. \nonumber \\
\label{A24}
\end{eqnarray}

\acknowledgments
This work is supported by the Natural Science Foundation of China under the Grant No. 12065006. \\

\providecommand{\href}[2]{#2}\begingroup\raggedright

\end{document}